\documentclass[nofootinbib,prl,superscriptaddress,floatfix,twocolumn]{revtex4-2}

\usepackage{graphics}
\usepackage{graphicx}
\usepackage{bbding}
\usepackage{subfigure}
\usepackage{mathrsfs}
\usepackage{amsmath}
\usepackage[table,xcdraw]{xcolor}
\usepackage{color}
\usepackage{natbib}
\usepackage{bookmark}

\begin{document}
\date{\today}

\title{L\'evy Walks and Path Chaos in the Dispersal of Elongated Structures Moving across Cellular Vortical Flows}

\author{Shi-Yuan Hu}
\affiliation{Applied Mathematics Lab, Courant Institute of Mathematical Sciences, New York University, New York, NY 10012, USA}
\affiliation{Department of Physics, New York University, New York, NY 10003, USA}
%%%%%%%
\author{Jun-Jun Chu}
\affiliation{School of Physics Science and Engineering, Tongji University, Shanghai 200092, China}
%%%%%%%
\author{Michael J. Shelley}
\email[]{shelley@cims.nyu.edu}
\affiliation{Applied Mathematics Lab, Courant Institute of Mathematical Sciences, New York University, New York, NY 10012, USA}
\affiliation{Center for Computational Biology, Flatiron Institute, New York, NY 10010, USA}
%%%%%%%
\author{Jun Zhang}
\email[]{jun@cims.nyu.edu}
\affiliation{Applied Mathematics Lab, Courant Institute of Mathematical Sciences, New York University, New York, NY 10012, USA}
\affiliation{Department of Physics, New York University, New York, NY 10003, USA}
\affiliation{NYU-ECNU Institute of Physics at NYU Shanghai, Shanghai 200062, China}

%%%%%%%%%%%%%%%%%%%%%%
\begin{abstract}
%%%%%%%%%%%%%%%%%%%%%%
In cellular vortical flows, namely arrays of counter-rotating vortices, short but flexible filaments can show simple random walks through their stretch-coil interactions with flow stagnation points. Here, we study the dynamics of semi-rigid filaments long enough to broadly sample the vortical field. Using simulation, we find a surprising variety of long-time transport behavior -- random walks, ballistic transport, and trapping -- depending upon the filament's relative length and effective flexibility. Moreover, we find that filaments execute L\'evy walks whose diffusion exponents generally decrease with increasing filament length, until transitioning to Brownian walks. Lyapunov exponents likewise increase with length. Even completely rigid filaments, whose dynamics is finite-dimensional, show a surprising variety of transport states and chaos. Fast filament dispersal is related to an underlying geometry of ``conveyor belts''. Evidence for these various transport states are found in experiments using arrays of counter-rotating rollers, immersed in a fluid and transporting a flexible ribbon.
\end{abstract}

\maketitle
%%%%%%%%%%%%%%%%%%%%%%%%%%%%%%%%%%%%%%%%%%%%%%%%%%%%%%%
Flows at low Reynolds number (Re) are typically laminar and regular. However, chaotic or turbulent dynamics can emerge, such as by flowing through complex geometries~\cite{Stroock}, adding elastic polymers~\cite{Groisman00,Arratia,Steinberg}, and exploiting the hydrodynamic interactions between suspended passive particles in externally driven flows~\cite{Pine}, or between active motile ones~\cite{DombrowskiEtAl2004,SS2013}. The understanding of how random and complex dynamics can emerge in simple flows at low Re is important in numerous applications~\cite{Groisman01,Stone,Lee1,Ward,Nam,Yazdi,Karimi,Lu}.

Time-independent cellular vortical flows are simple flows with inherent characteristic scales and closed streamlines. They often arise as simplified models for flows in nature~\cite{Stommel,Solomon88,Ariel17} and have been realized in different experiments~\cite{Rothstein,Ouellette,Wandersman}. Tracer particles in such flows simply follow closed streamlines. When the particle size is negligible compared with the characteristic flow scale, complex dynamics, such as aggregation~\cite{Torney} and L\'evy walks~\cite{Ariel15,Ariel17,Ariel20}, have been found for active particles. For passive flexible filaments, which can show complex deformations even in simple shear flow~\cite{LaGrone}, complex dynamics arises differently: driven by buckling instabilities near the flow stagnation points, the filaments behave as Brownian walkers across the array~\cite{Young,Wandersman,Quennouz}. The recent literature on filament dynamics is reviewed in Ref.~\cite{Roure}.

For finite-extent filaments, transport is determined by flows sampled nonlocally along the filament, in contrast to small or compact particles. Complex and different behaviors may be generated and controlled through the coupling of flows and filaments at different length scales. In this Letter, we use experiments and a comprehensive set of numerical simulations to investigate the transport of rigid and semi-rigid filaments when their length $L$ is comparable to the vortex size $W$ in an idealized Stokesian cellular flow. In this regime, the background vortices can be viewed as `soft' scatters for the filaments and the dynamics shows similarity to billiards systems~\cite{Zarfaty18,Zarfaty19,Klages}. We construct a phase diagram that shows the rich variety of transport states possible for filaments moving across this simple low-Re flow. In particular, we find that as $L/W$ increases, there exists a transition through L\'evy walks, of generally decreasing diffusion exponents and increasing Lyapunov exponents, to Brownian walks. L\'evy walks have been found in the dynamics of active elongated particles in cellular flows when $L/W\ll 1$, but arising there due to particle motility~\cite{Ariel17}. Quite remarkably, even completely rigid filaments, described by only center-of-mass (CoM) position and orientation, show these varieties of random walks and chaotic motions. 

Besides $L$ and $W$, the dynamics of a flexible filament depends also upon its elastohydrodynamic length $l_{e} \sim \left(B\big/\mu U_0\right)^{1/3}$~\cite{Wiggins}, where $B$ is filament rigidity, $\mu$ is fluid viscosity, and $U_0$ is the characteristic flow velocity. The interplay among the three length scales are captured by two dimensionless control parameters: the relative length $\gamma=L/W$ and the effective flexibility $\eta \propto \left(L/l_{e}\right)^{3}$. The filament appears to be more `flexible' when $\eta$ is larger.

\textit{Motivating Experiment}.---We set up a cellular flow structure by immersing a square 9-by-9 roller array into a tank of pure glycerol (see Fig.~\ref{fig1} and details in supplemental material~\cite{sm}). The rollers are interconnected and driven by a stepper motor. Through viscous coupling each roller rotates the fluid around it, with nearest neighbor rollers being counter-rotating. Flexible ribbons, made from audio tape, are transported in the cellular flow and stay right beneath the fluid surface, resembling 2D motions. The effective flexibility $\eta \approx$ 7.5--354. The $\text{Re} = \rho U_0 W/\mu \approx$ 0.1--1, where $\rho$ is the density of glycerol. As shown in Fig.~\ref{fig1}, a few interesting patterns of ribbon's motion are identified. For moderate $\gamma$, despite occasional trappings, the ribbons can reach the edge of the cellular flow through undulating steps directed along diagonals or the $\pm x$ or $\pm y$ directions [Figs.~\ref{fig1}(a)--(c)]. However, for larger $\gamma$, the ribbons may meander around for a long time and make many turns before getting to the edge [Fig.~\ref{fig1}(d)]. When $\eta$ is sufficiently large, the ribbons are bent with large deformation and often trapped inside one of the fluid vortices.
%%%
\begin{figure}[t]
\centering
\includegraphics[bb= 0 15 365 267, scale=0.67,draft=false]{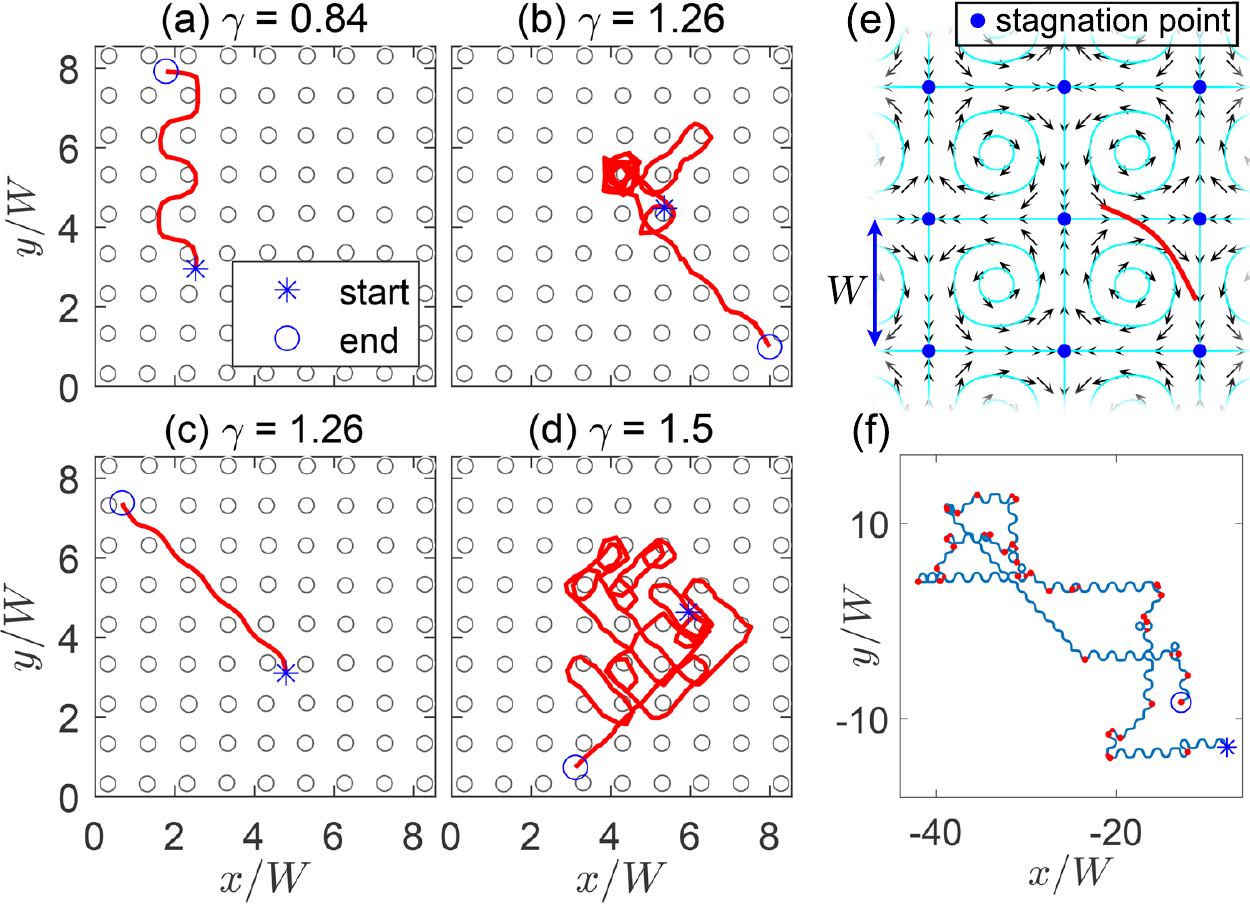}
\caption{Experimental CoM trajectories of flexible ribbons in a 9-by-9 cellular flow, as shown on the left. Circles indicate the locations of the spinning rollers. (a) A strongly undulating step along $+y$ direction with $\eta \approx 25$. (b) A meandering trajectory with a long undulating diagonal step with $\eta \approx 85$. (c) Diagonal step with $\eta \approx 85$, (d) Meandering trajectory with many turns with $\eta \approx 118$. (e) Snapshot from a numerical simulation with $\eta=100$ and $\gamma=1$, showing a flexible filament moving in the cellular flow described by Eq.~(\ref{eq1}). Black arrows are the background cellular flow and cyan closed curves are the streamlines. (f) A typical trajectory from simulation with $\gamma = 1$ and $\eta=0.5$. The red points are turning points separating two different steps.}
\label{fig1}
\end{figure}
%%%

\textit{Simulation and Model}.---As a model, we consider slender, inextensible, and elastic filaments of radius $R$ and length $L$ (with aspect ratio $\epsilon = R/L \ll 1$) moving in a Stokesian flow. Lengths are scaled on $L$, velocity on $U_0$, and time on $L/U_0$. The stream function $\Phi_\gamma$ of the background flow $\textbf{U}$ is given by
\begin{equation}\label{eq1}
\Phi_\gamma = (\pi\gamma)^{-1}\sin(\pi\gamma x)\sin(\pi\gamma y),
\end{equation}
which has stagnation points at $(n,m)\gamma^{-1}$ for $n,m$ integers. The unit periodic cell is composed of four counter-rotating vortices [Fig.~\ref{fig1}(e)]. The filament centerline, denoted $\textbf{r}(s,t)$, is parametrized by a signed arclength $s\in[-1/2,1/2]$. From the leading-order slender body approximation~\cite{KR1976}, the centerline velocity $\textbf{r}_{t}$ is governed by a local balance of drag force with the filament force (per unit length) upon the fluid,
%%%
\begin{equation}\label{eq2}
\eta \left(\textbf{I}-\textbf{r}_{s}\textbf{r}_{s}/2\right) (\textbf{r}_{t}-\textbf{U}[\textbf{r}]) = -\textbf{r}_{ssss}+(T\textbf{r}_{s})_s,
\end{equation}
%%%
where the effective flexibility $\eta = 8\pi\mu U_0 L^3 \big/cB$ with $c=|\ln(\epsilon^2 e)|$ and $\textbf{U}[\textbf{r}]$ is the background flow along the filament centerline. The tensor $\textbf{I}-\textbf{r}_s\textbf{r}_s/2$ captures the drag anisotropy of the filament. The filament force is described by Euler-Bernoulli elasticity: $\textbf{f} = \textbf{r}_{ssss}-(T\textbf{r}_{s})_{s}$, where determination of the tension $T$ enforces filament inextensibility. Equation~(\ref{eq2}) is evolved numerically using a second-order finite difference method and implicit time-stepping, while imposing zero-force and -torque boundary conditions~\cite{sm}.

%%%
\begin{figure}[b]
\centering
\includegraphics[bb=0 15 360 242, scale=0.66,draft=false]{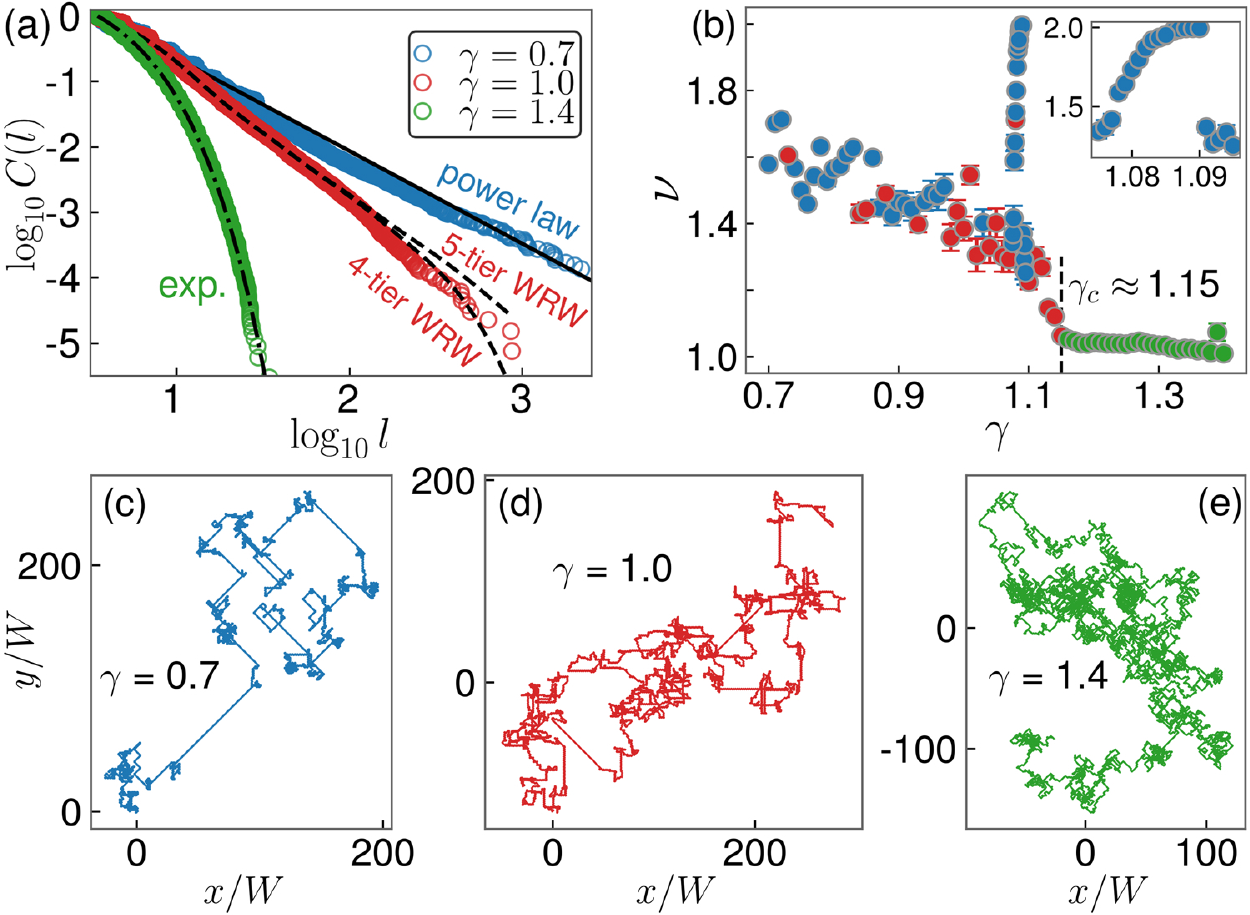}
\caption{From L\'evy walks to Brownian walks for rigid filaments. Blue: power-law $\phi(l)$, red: WRWs, and green: exponential $\phi(l)$. (a) $C(l)$ for $\gamma = 0.7$, 1.0, and 1.4, with best-fit distributions. With more statistics, higher-order terms in WRWs are needed~\cite{sm}. (b) $\nu$ as a function of $\gamma$. Error bars represent uncertainties due to initial conditions. Inset shows $\nu$ versus $\gamma$ for $1.075 \le \gamma \le 1.095$. (c)--(e) Typical trajectories corresponding to the three cases shown in (a).}
\label{fig2}
\end{figure}
%%%
The case of a rigid straight filament is informative~\cite{sm}. We take $\textbf{r}(s,t) = \textbf{r}_c(t)+s\hat{\textbf{p}}(t)$, where $\textbf{r}_c$ is the CoM position and $\hat{\textbf{p}} = (\cos\theta_c,\sin\theta_c)$ with $\theta_c$ the filament orientation. Under zero-force and -torque conditions, the equations of motion are purely kinematic:
\begin{eqnarray}
  \dot{\textbf{r}}_c &=& -\nabla^{\perp}_{\textbf{r}_c}\mathcal{H}
  \text{ with  }
  \mathcal{H} = \int_{-1/2}^{1/2} \Phi_\gamma[\textbf{r}(s,t)]\,ds,
  \label{eq3} \\
  \dot{\theta}_c &=& 12\mathcal{H}
  -6\left(\Phi_\gamma\left[\textbf{r}(1/2,t)\right]
  +\Phi_\gamma\left[\textbf{r}(-1/2,t)\right]\right),
  \label{eq4}
\end{eqnarray}
where $\nabla^{\perp}=(-\partial_y,\partial_x)$. In the ``point limit'' $\gamma=0$, particle transport [Eq.~(\ref{eq3})] is Hamiltonian, local, and decoupled from particle rotational dynamics. Increasing $\gamma$ increases the averaging (over particle length) of the background flow, while also increasing the coupling of particle translation to rotation. This increase in system dimension leads to loss of integrability and allows for transport chaos. Equations~(\ref{eq3}) and (\ref{eq4}) are evolved using a fourth-order Runge-Kutta scheme. Our simulations typically run for $2\times 10^4\ L/U_0$ to capture long-time dynamics.

\textit{Rigid filament simulations}.---The filament CoM trajectories show strong dependence on $\gamma$ and the initial conditions. A common statistical measure of a complex trajectory is its step-length distribution $\phi(l)$~\cite{Klafter}. The step-length $l$ is the straight-line distance between successive turning points, which separate two steps along different directions (see Fig.~\ref{fig1}(f) and Ref.~\cite{sm}). For rigid filaments of different $\gamma$, Fig.~\ref{fig2}(a) shows the complement of the cumulative distribution, $C(l) = 1-\int_{a}^{l}\phi(l)dl$, for $l$ in 400 trajectories with random initial conditions. We fit several random walk models using a maximum likelihood method~\cite{Edwards,Newman}. At small $\gamma$, the best-fit models are mostly power laws given by $\phi(l) \propto l^{-(1+\beta)}$ with $0<\beta<2$, indicating L\'evy walks~\cite{Klafter}. The trajectories show clusters of short steps interspersed with long steps [Fig.~\ref{fig2}(c)]. At large $\gamma$, $\phi(l)$ fits well to exponential functions and the filaments display Brownian walks [Fig.~\ref{fig2}(e)]. At intermediate values of $\gamma$, with the exception of $\gamma$ around 1.08--1.09, the best-fit models are mostly Weierstrassian random walks (WRW), which have been found in the studies of animal search strategies and random walks in bacterial swarms~\cite{Hughes,Andy,Andy2,Ariel17}. The $\phi(l)$ of the WRW is given by a hierarchical sum of exponential distributions with mean $b^{-(j+1)}$ weighted by $q^{-(j+1)}$: $\phi(l) \propto \sum_{j=0}^{J} q^{-(j+1)}b^{j+1}\exp\left(-b^{j+1}l\right)$, which resembles a power law when $J\rightarrow\infty$ and degenerates into an exponential distribution when $J = 0$. 

\begin{figure}[b]
\centering
\includegraphics[bb = 0 20 358 225, scale=0.66,draft=false]{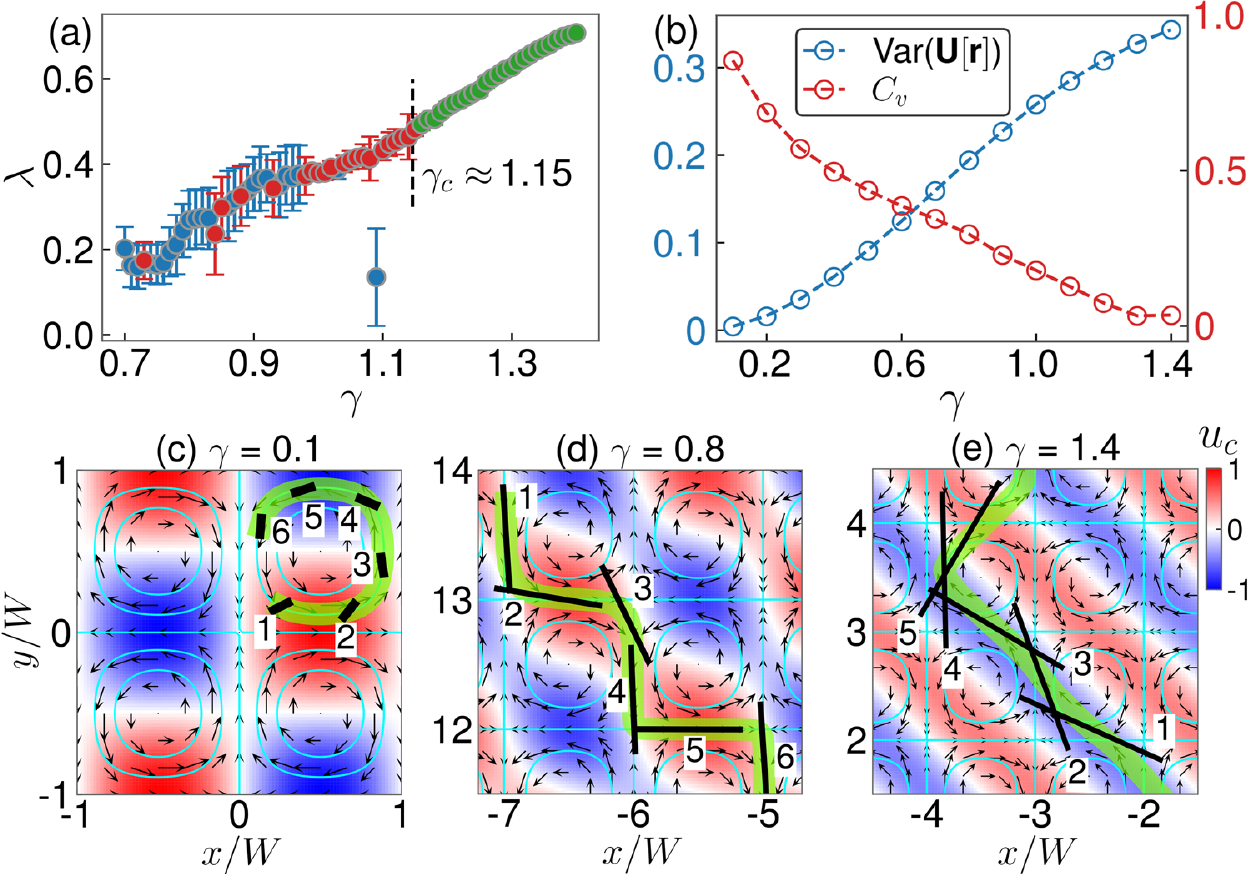}
\caption{(a) Lyapunov exponent $\lambda$ as functions of $\gamma$ for rigid filament. The large error bars for L\'evy walks are due to the nonuniformity of the phase space. (b) $\text{Var}(\textbf{U}[\textbf{r}])$ (left) and $C_{v}$ (right) as a function of $\gamma$. (c), (d), and (e) Maps of positive $u_c$ (red, rightward motion) and negative $u_c$ (blue, leftward motion) computed with $\theta_c=3\pi/4$ (see text for detail). Snapshots of the motion of rigid filaments are also shown with the time ordering labeled by numbers. Green thick lines trace the CoM trajectories.}
\label{fig3}
\end{figure}

The above transition through L\'evy walks to Brownian walks is confirmed from the MSD scaling exponent~\cite{Ralf}, $\big\langle \delta^{2}(\tau) \big\rangle \sim \tau^{\nu}$ [Fig.~\ref{fig2}(b)]. In general, $\nu$ decreases as $\gamma$ increases. The critical $\gamma$ separating the two transport behaviors is around 1.15. At $\gamma$ around 1.08--1.09 [Fig.~\ref{fig2}(b) inset], nearly ballistic trajectories are observed with $\nu \approx 2$. This is due to a geometric match of the filament length to the flow periodicity and unstable under small perturbations of the flow field~\cite{sm}. As shown in Fig.~\ref{fig3}(a), the transition is accompanied with a growth in the Lyapunov exponent $\lambda$~\cite{sm,Benettin,Ramasubramanian}: Brownian walks are more chaotic than L\'evy walks. The phase space shows complex structures with strong dependence on $\gamma$, and L\'evy walks are related to particle stickiness to regular islands~\cite{sm}. A transition from L\'evy walks to Brownian walks has been found in the transport of ions in optical lattice~\cite{Marksteiner,Katori}. The presence of L\'evy walks is also known in Hamiltonian chaos, but the transitions are typically abrupt~\cite{Chernikov,Zaslavsky93,Benkadda,Harsoula}.
%%%%%%%%%%%%
%%%%%%%%

\textit{Mechanism}.---We attribute the emergence of chaos and different random walks to the nonlocal geometrical averaging of the background flow by the filament from its broad extension across vortices. Consider rigid filament, from Eq.~(\ref{eq3}), the CoM velocity of the filament is $\textbf{v}_c=(u_c,v_c)=\int_{-1/2}^{1/2} \textbf{U}[\textbf{r}(s)]\,ds$. We first compute the variance of $\textbf{U}[\textbf{r}(s)]$: $\text{Var}\left(\textbf{U}[\textbf{r}]\right) =\big\langle\int_{-\frac{1}{2}}^{\frac{1}{2}}\left(\textbf{U}[\textbf{r}(s)]-\textbf{v}_c\right)^2\,ds \big\rangle$, where the average is taken with respect to $\textbf{r}_c$ over the entire unit cell and $\theta_c$ over [0, $2\pi$). With the increase of $\gamma$ [Fig.~\ref{fig3}(b) left], $\text{Var}(\textbf{U}[\textbf{r}])$ becomes larger, i.e., the background flow that the filament experienced on average becomes more variable. We also compute the correlation function between the unit velocity vectors of filament's two ends ($s=1/2,-1/2$): $C_{v} = \langle \hat{\textbf{v}}(-1/2)\cdot \hat{\textbf{v}}(1/2)\rangle$, where the average is along CoM trajectories [Fig.~\ref{fig3}(b) right]. When $C_v$ is large, the filament is likely to be translated along a flow but to be turned around when $C_v$ is small. Both $\text{Var}(\textbf{U}[\textbf{r}])$ (increasing with $\gamma$) and $C_v$ (decreasing with $\gamma$) show that long filaments with large $\gamma$ can hardly travel long unidirectional steps but rather turn around and take seemingly random and diffusive motions. On the other extreme at very small $\gamma$, filaments are too short to perceive any flows outside the local circulation within which it resides and most of the filaments are trapped except those initially close to the separatrices. Therefore, the only possible long-distance travelers are those filaments of intermediate lengths. 

Indeed, we find some clues by mapping the spatial distributions of $x$-component CoM velocity $u_c(\textbf{r}_c,\theta_c)$. In Figs.~\ref{fig3}(c)--\ref{fig3}(e), the maps of $u_c$ were made with $\theta_c=3\pi/4$, but any other values of $\theta_c$ in the second and fourth quadrants would result in similar maps but slightly smaller magnitude of $u_c$. When $\gamma \ll 1$ [Fig.~\ref{fig3}(c)], areas of positive $u_c$ values (red, rightward motion) and negative $u_c$ values (blue, leftward motion) are isolated from each other, filament cannot travel across vortices. As $\gamma$ increases, the red and blue regions start to deform and morph into many alternating `conveyor belts' flowing towards opposite directions [Figs.~\ref{fig3}(d), \ref{fig3}(e)]. If $\theta_c$ lies in the first and third quadrants, the `conveyor belts' will take the other diagonals oriented $\pi/2$ from those in Figs.~\ref{fig3}(d) and \ref{fig3}(e). The filament can now move across vortices and travel long steps. We see two competing effects at work as $\gamma$ further increases: longer filament promotes the formation of `conveyor belts' but at the same time it is more likely to turn and change directions. The latter effect is demonstrated in Fig.~\ref{fig3}(e) as a long filament is captured first by an opposite `conveyor belt' (label 4) and then turns (label 5) with its two ends moving oppositely.

%%%
\begin{figure}[b]   
\centering
\includegraphics[bb=0 15 365 205, scale = 0.67,draft=false]{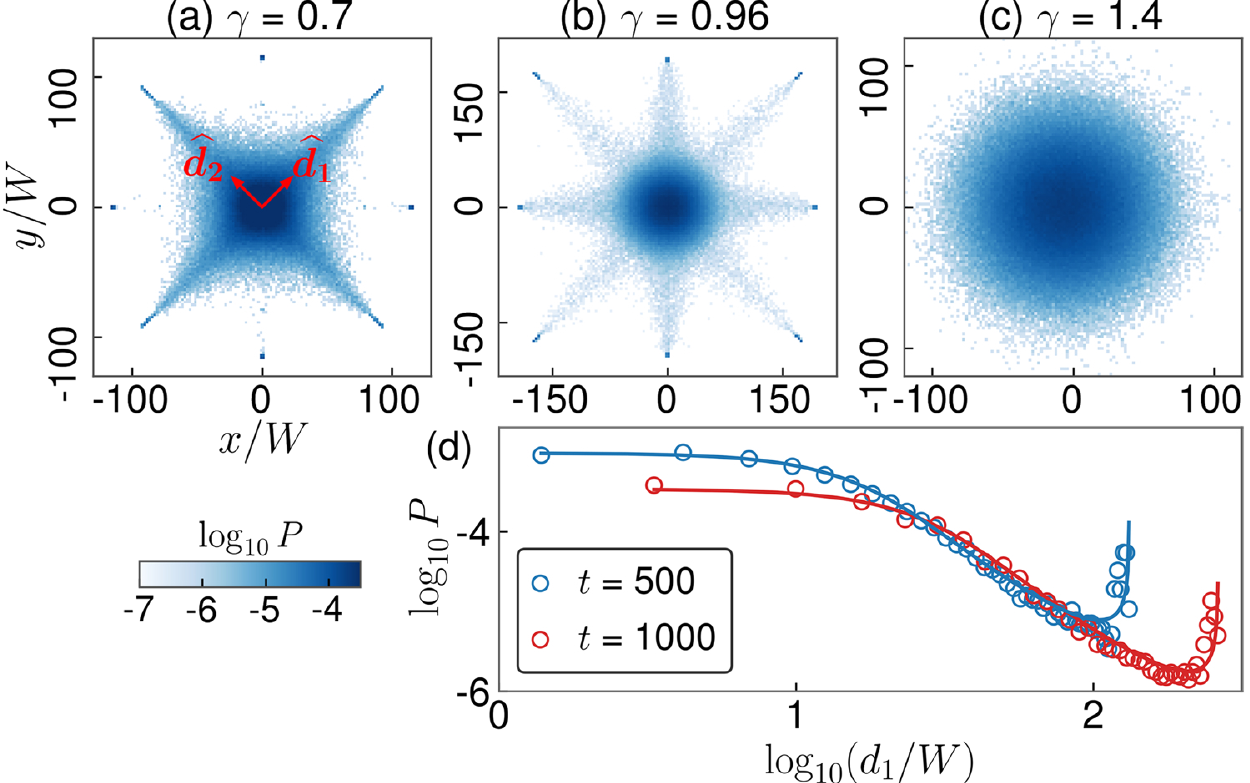}
\caption{Dispersal patterns of $10^6$ rigid filaments. Probability density function (PDF) $P(x,y)$ at $t=500$ for (a) $\gamma=0.7$, (b) $\gamma=0.96$, and (c) $\gamma=1.4$. (d) PDF along direction $\hat{d}_1$ shown in (a) at two different time instants for $\gamma=0.7$. Solid lines are the fittings of the theoretical result given by Eq.~(\ref{eq5}).}
\label{fig4}
\end{figure}
%%%
The patterns of filament dispersal at scales much larger than $W$ are significantly different for different random walks. For those performing L\'evy walks they are strikingly anisotropic. Figure~\ref{fig4}(a) shows that for $\gamma =0.7$, the probability density function (PDF) $P(\textbf{x},t)$ of finding a filament at position $\textbf{x}$ at time $t$ after starting off with random initial conditions from the unit cell centered at the origin has a `{\color{darkgray}\reflectbox{\rotatebox[origin=c]{45}{\scriptsize\FourStar}}}'-like structure with four branches extending along the diagonals specified by $(\hat{d}_1, \hat{d}_2)$. Such anisotropy arises from the long unidirectional diagonal steps due to the `conveyor belts'. In particular, a sharp peak exists at the far front of each branch [Fig.~\ref{fig4}(b)] reflecting the microscopic geometry of the L\'evy walks~\cite{Zaburdaev}: filaments can only move along $\hat{d}_1$ or $\hat{d}_2$ at each step. The PDF along $\hat{d}_1$ is given by a product of a 1D L\'evy distribution and a pre-factor that accounts for the decrease in the spread of the PDF along $\hat{d}_2$~\cite{Zaburdaev},
%%%%%
\begin{equation}\label{eq5}
P(d_{1},t) \propto \left(1-d_{1}/(ct)\right)^{-1/\beta} \mathcal{L}^{\sigma}_{\beta}(d_{1}),
\end{equation}
%%%%%
where $c$ is the average speed of the filaments and $\mathcal{L}_{\beta}^{\sigma}$ is a 1D L\'evy distribution with exponent $\beta$ and scale parameter $\sigma$ ($\propto t^{1/\beta}$). Equation~(\ref{eq5}) agrees well with the simulation result [Fig.~\ref{fig4}(d)]. As $\gamma$ increases, undulating steps along $\pm\,x$ or $\pm\,y$ directions become more frequent and 8-fold `{\color{darkgray}{\footnotesize\EightStarTaper}}'-like patterns are formed [Fig.~\ref{fig4}(b)]. While almost unapparent, these additional branches can be faintly discerned in Fig.~\ref{fig4}(a). Eventually in the Brownian-walk regime for sufficient large $\gamma$, in sharp contrast to L\'evy walks, the PDF follows an isotropic 2D Gaussian distribution with its variance scaling linearly with time $t$ [Fig.~\ref{fig4}(c)].

%%%
\begin{figure}[b]
\centering
\includegraphics[bb=5 15 350 220, scale = 0.66,draft=false]{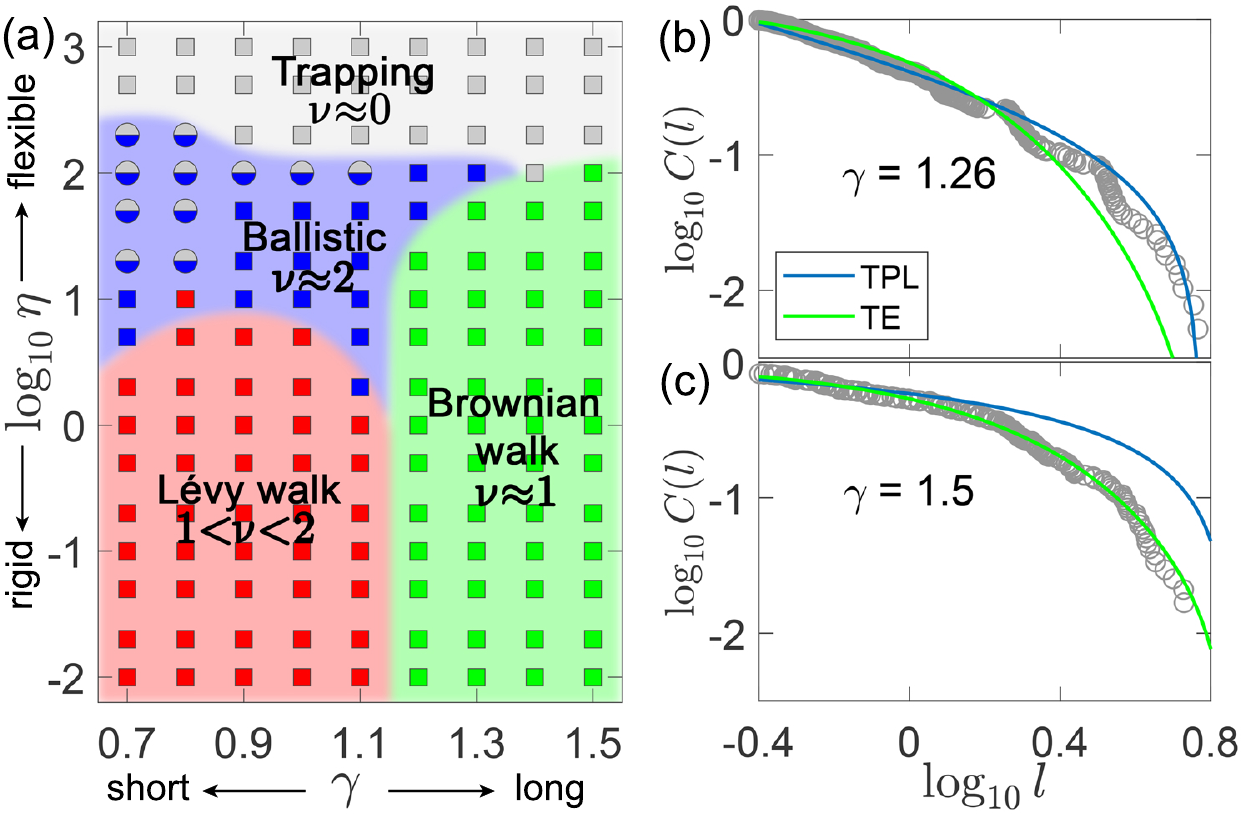}
\caption{(a) From simulation, the phase diagram of the transport states constructed using MSD. Depending on the initial conditions, ballistic states may coexist with trapping states (circles). (b), (c) $C(l)$ from 52 trajectories in experiments for two different values of $\gamma$.}
\label{fig5}
\end{figure}
%%%
\textit{Flexible filament simulations}.---By extensively surveying the phase space of $\gamma$ and $\eta$, for the first time, we construct a phase diagram showing various transport states (see Fig.~\ref{fig5}(a) and Ref.~\cite{sm}). Highly flexible filaments ($\eta \gtrsim 10^{2}$) are deformed and trapped inside vortices for all values of $\gamma$, as is also observed in the experiments. For relatively rigid filaments ($\eta \lesssim 10$), the transport states are determined by $\gamma$. The boundary that separates L\'evy walks ($1<\nu<2$) and Brownian walks ($\nu \approx 1$) is located around $\gamma = 1.15$. At intermediate $\eta$ between $10-10^2$, the filament first meanders around for a short period before moving indefinitely along diagonals or $\pm$x and $\pm$y directions. We call this type of transport behavior a ballistic state with $\nu \approx 2$. Even within the ballistic state, the filament dispersal patterns show strong dependence on both $\gamma$ and $\eta$ and filaments with different lengths and flexibility are dynamically sorted~\cite{sm}. The dispersal rate of filaments is largest in ballistic state, followed by L\'evy walks and then Brownian walks. There is no long-range transport in the trapping state. 

\textit{Experimental evidence}.---We find evidence that shorter ribbons perform L\'evy walks while longer ribbons perform Brownian walks in the experiments~\cite{sm}. Figures~\ref{fig5}(b) and \ref{fig5}(c) show two examples. For $\gamma = 1.26$, the best-fit model is truncated power law (TPL)~\cite{Raichlen} supporting a L\'evy walk; for $\gamma = 1.5$, the step length distribution better resembles a truncated exponential (TE) model supporting a Brownian walk. The $\gamma$ value that separates the two states is estimated to be around $1.4$, which differs from the value found in the simulation. This discrepancy is possibly caused by the presence of the roller boundaries. Despite the subtle difference in $C(l)$ due to the limited size of the flow field in the experiments, the trajectories for the two states are significantly different from each other as depicted in Fig.~\ref{fig1}: with more turns in the trajectories, longer ribbons take much longer time to reach the edge of the flow field than shorter ribbons. Our experiments on the effect of $\eta$ are limited, but for large $\eta$, the ribbon is typically bent around one of the rollers and trapped for long time.

\textit{Discussion}.---Our experiments and simulations demonstrate that this simple system of a semi-rigid filaments moving in a Stokesian cellular vortical flows has a surprisingly rich range of dispersal dynamics, including L\'evy walks and path chaos. For $\gamma\gg 1$ our limited simulations show mostly Brownian walks due to averaging over multiple vortices. The emergence of cross-vortex motion and chaos in our system does not require the flow itself to be time-dependent and chaotic as it does for tracer particles~\cite{Solomon88,Solomon93}. It arises from the elongated body being able to broadly sample the background vortical field, and the strong coupling of rotational to translational dynamics afforded by that elongation. The cross-streamline motion and escape from local flows shown by the semi-rigid filaments have implications for efficient fluid mixing at low Re by additives~\cite{Groisman01, Lee2}. Similar to the billiards system~\cite{Zarfaty18,Zarfaty19}, the anisotropic dispersals of filaments in the L\'evy-walk state and ballistic state are also originated from long unidirectional steps pre-programmed by the fundamental geometries of the backgrounds. However, the dominant directions of motion depend on the relative length $\gamma$ and effective flexibility $\eta$ of filaments in our system but are fully specified by the geometries of scatterers in the billiards. Most prominently, various transport states can be achieved by tuning different length scales, which also serves as the underlying mechanism of gel electrophoresis~\cite{Dorfman}. Our results may open up new possibilities for efficient dynamical sorting of elongated particles and semi-flexible biopolymers~\cite{Huang,Dorfman,Sajeesh,Brato}.

%%%%%%%%%%%%%%%%%%%%%%%%%%%%%%%%
\bigskip
We thank L. Ristroph, E.A. Spiegel, Y.-N. Young, and J.-Q. Zhong for inspiring questions and helpful discussions. We also thank the anonymous reviewers for their inciteful criticisms and suggestions. S.Y.H. gratefully acknowledge the MacCracken Fellowship provided by New York University. M.J.S. and J.Z. acknowledge support from National Science Grant CBET-1805506, and J.Z. acknowledges support from NSFC-11472106.
%%%%%%%%%%%%%%%%%%%%%%%%%%%%%%%%
%%%%%%%%%%%%%%%%%%%%%%%%%%%%%%%%%%%%%%%%%

\end{document}

\date{\today}
\title{Supplemental Material for \\
L\'evy Walks and Path Chaos in the Dispersal of Elongated Structures Moving across Cellular Vortical Flows}

\author{Shi-Yuan Hu}
\affiliation{Applied Mathematics Lab, Courant Institute of Mathematical Sciences, New York University, New York, NY 10012, USA}
\affiliation{Department of Physics, New York University, New York, NY 10003, USA}
\author{Jun-Jun Chu}
\affiliation{School of Physics Science and Engineering, Tongji University, Shanghai 200092, China}
\author{Michael Shelley}
\email[]{shelley@cims.nyu.edu}
\affiliation{Applied Mathematics Lab, Courant Institute of Mathematical Sciences, New York University, New York, NY 10012, USA}
\affiliation{Center for Computational Biology, Flatiron Institute, New York, NY 10010, USA}
\author{Jun Zhang}
\email[]{jun@cims.nyu.edu}
\affiliation{Applied Mathematics Lab, Courant Institute of Mathematical Sciences, New York University, New York, NY 10012, USA}
\affiliation{Department of Physics, New York University, New York, NY 10003, USA}
\affiliation{New York University-East China Normal University Institute of Physics, New York University Shanghai, Shanghai 200062, China}

\maketitle

%%%%%%%%%%%%%%%%%%%%%%%%%%%%%%%%
\section{V\lowercase{ideo} I\lowercase{nformation}}
%%%%%%%%%%%%%%%%%%%%%%%%%%%%%%%%
Video1 (simulation): Filament in trapping state with $\gamma = 1.2$ and $\eta = 1000$.

Video2 (simulation): Filament in ballistic state taking strongly undulating step along $x$ direction with $\gamma = 1$ and $\eta=50$.

Video3 (simulation): Filament in ballistic state taking undulating step along diagonal with $\gamma = 1.2$ and $\eta=50$.

Video4 (simulation): Filament in L\'evy-walk state with $\gamma = 0.9$ and $\eta = 0.1$.

Video5 (simulation): Filament in Brownian-walk state with $\gamma = 1.4$ and $\eta = 0.5$.

Video6 (experiment): Directional motion of ribbon with $\gamma = 1.26$ and $\eta \approx 85$. (Trajectory Duration 30.5 sec.)

Video7 (experiment): Meandering motion of ribbon with $\gamma = 1.5$ and $\eta \approx 118$. (Trajectory Duration 204 sec.)
Occasionally, the experimental center-of-mass (CoM) trajectory lacks smoothness. This is largely due to the imperfect tracking caused by the extremely thin ribbon. When our tracking algorithm processes the video recordings (with a broad view field), it fails at times to recognize the entire filament.

%%%%%%%%%%%%%%%%%%%%%%%%%%%%%%%%
\section{E\lowercase{xperimental} D\lowercase{etails}}
%%%%%%%%%%%%%%%%%%%%%%%%%%%%%%%%
%%%%%%
\begin{figure*}[h]
\centering
\includegraphics[bb=0 0 340 170, scale=0.8,draft=false]{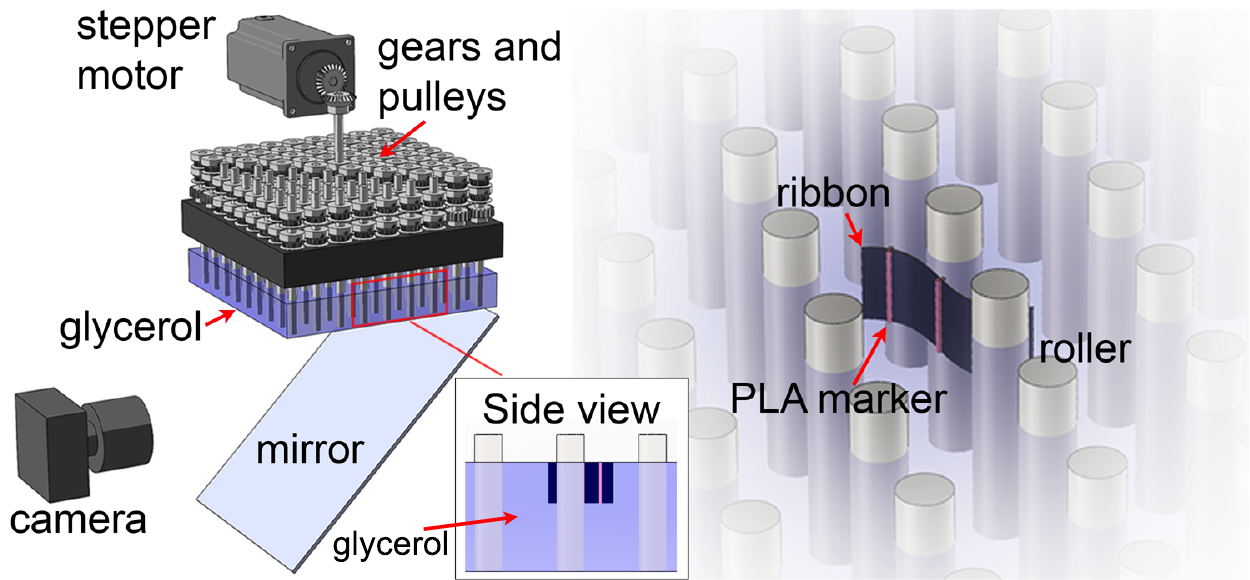}
\caption{Schematic of the experimental setup as shown on the left. Right panel details a ribbon moving between rollers.}
\label{fig_a1}
\end{figure*}
%%%%

The experiment setup is shown in Fig.~\ref{fig_a1}. We immerse a square 9-by-9 roller array into a tank of pure glycerol from the above. The fluid surface is free and its bottom is no-slip. The center-to-center distance between the nearest neighbor rollers is $W=19$ mm, and the diameter of the roller is 6.35 mm. Other parameters are listed in Table~\ref{table_a1}. The rollers are inter-connected with gears, pulleys, and timing belts and driven by a stepper motor, such that nearest neighbor rollers are counter-rotating and rollers in the same diagonal are co-rotating. Each roller rotates the fluid around it through viscous coupling and a cellular flow structure is formed. Different from our method, another way to form cellular flows is using electromagnetic forcing~\cite{Wandersman}.

In our experiments, we transport flexible ribbons made from audio tapes that are slightly denser than glycerol. To prevent the ribbon from sinking, we attach equally-spaced PLA (polylactic acid) markers, thin rods of 1 mm diameter, to the ribbon because PLA is lighter than glycerol. As a result, the ribbon stays right beneath the free surface and remains vertical due to the quasi-2D flow. Since the ribbon width is small compared with the depth of the fluid, we can neglect the effect of the no-slip bottom of the tank on ribbon's motion. Therefore, the motion of the ribbon is essentially 2D. The PLA markers can also make the video tracking easier. The motion of the ribbon is recorded by a camera through a reflection mirror at $45^{\circ}$ with respect to the horizontal plane. We then process the video using Matlab Image Processing Toolbox and obtain the CoM trajectories. 
In the experiment, the two control parameters are the rotation frequency $f_{0}$ of the stepper motor and the ribbon length $L$. We vary $f_{0}$ mostly between 0.3 Hz and 1.5 Hz and experiment on three values of $L$: 16 mm, 24 mm, and 28.5 mm. The effective flexibility $\eta$ can be estimated. For example, when $L$ = 24 mm and $f_{0}$ = 1 Hz,
\begin{equation}\label{eq.a1}
    \eta = \frac{8\pi\mu U_0 L^3}{cB} = \frac{8\pi\mu f_0 L^4}{cB} \approx 121,
\end{equation}
where $\mu$ is the dynamical viscosity of glycerol and the characteristic flow velocity $U_0 = f_0 L$.
%%%%
\begin{table}[h]
    \centering
    \begin{tabular}{p{5.5cm}p{3.8cm}}
        \multicolumn{2}{c}{Geometric parameters} \\ 
        \hline
        distance of neighbor rollers $W$  & 19 mm   \\ 
        roller diameters  &    6.35 mm  \\ 
        fluid depth & 25 mm \\
        tank size & 171 mm$\times$171 mm$\times$30 mm \\
        \hline \\[0.1mm]
        \multicolumn{2}{c}{Material parameters}  \\
        \hline 
        ribbon width   &  6$\times 10^{-3}$ m  \\ 
        ribbon thickness   &  2.54$\times 10^{-5}$ m \\
        ribbon bending rigidity $B$  &  7.6$\times 10^{-9}$ J$\cdot$m \\
        ribbon density   &   1.3--1.4$\times 10^{3}$ kg/m$^{3}$ \\ 
        glycerol density   &  1.26$\times 10^{3}$ kg/m$^{3}$ \\ 
        glycerol dynamic viscosity, $\mu$ & 1.412 Pa$\cdot$s \\ 
        PLA density  &  1.24$\times 10^{3}$ kg/m$^{3}$ \\ 
        diameter of PLA marker  &  1$\times 10^{-3}$ m \\ 
        \hline
    \end{tabular}
    \caption{Parameters of the experimental setup.}
    \label{table_a1}
\end{table}
%%%%%

%%%%%%%%%%%%%%%%%
\section{F\lowercase{ilament} D\lowercase{ynamics}}
%%%%%%%%%%%%%%%%%
In this section, we derive the equations [Eqs. (2)--(4) main text] governing the dynamics of slender and inextensible filaments moving in Stokesian flows. From a leading order slender body approximation, the perpendicular drag coefficient is twice the parallel drag coefficient and the filament velocity is described by, 
%%%
\begin{equation}\label{eq.a2}
8\pi\mu (\textbf{r}_t - \textbf{U}[\textbf{r}]) = c(\textbf{I}+\textbf{r}_s\textbf{r}_s)\textbf{f}.
\end{equation}
%%%
Here, $\textbf{U}$ is the background flow field, $\textbf{r}$ is the filament centerline parameterized by the signed arc length $s$, the tangent vector $\textbf{r}_s = (\cos\theta,\sin\theta)$ with $\theta$ the tangent angle, and $\textbf{r}_s\textbf{r}_s$ is the dyadic product. The filament force $\textbf{f}$ is described by Euler-Bernoulli elasticity: $\textbf{f} = -B\textbf{r}_{ssss}+(T\textbf{r}_s)_s$, where $B$ is the bending rigidity and $T$ is line tension used to enforce the inextensibility of the filament. Scale lengths on filament length $L$, velocities on the characteristic flow velocity $U_0$, and forces on $B/L^2$, we obtain Eq. (2) given in the main text,
%%%
\begin{equation}\label{eq.a3}
\eta (\textbf{r}_{t}-\textbf{U}[\textbf{r}])=\left(\textbf{I}+\textbf{r}_s\textbf{r}_s\right)\left(-\textbf{r}_{ssss}+(T\textbf{r}_s)_s\right),
\end{equation}
%%%
Filament tension $T$ can be determined by substituting the filament velocity $\textbf{r}_t$ into the inextensibility condition, $\textbf{r}_s\cdot \textbf{r}_{st} = 0$,
%%%
\begin{equation}\label{eq.a4}
2T_{ss}-\theta_{s}^{2}T = -\eta \textbf{U}_{s} \cdot \textbf{r}_s-6\theta_{ss}^{2}-7\theta_{s}\theta_{sss}+\theta_{s}^{4}.
\end{equation}
%%%
The tangent angle $\theta$ satisfies $\theta_t = \textbf{r}_{st} \cdot \textbf{r}_s^{\perp}$, where $\textbf{r}_s^{\perp} = (-\sin\theta, \cos\theta)$,
%%%
\begin{equation}\label{eq.a5}
\theta_{t} = \textbf{U}_{s}\cdot \textbf{r}_s^{\perp}+\eta^{-1}(-\theta_{ssss}+9\theta_s^2\theta_{ss}+3T_s \theta_s+T\theta_{ss}).
\end{equation}
%%%
As boundary conditions, we require force free and torque free conditions at both ends of the filament, which translate into,
%%%
\begin{equation}\label{eq.a6}
\begin{aligned}
\theta_s = 0, \theta_{ss} = 0,\ \text{and}\ T = 0,\ \text{at $s=-L/2$ and $s=L/2$}.
\end{aligned}
\end{equation}
%%%

For rigid filament, the centerline is described by $\textbf{r}(s)= \textbf{r}_c+s\hat{\textbf{p}}$, where $\textbf{r}_c$ is the CoM position and $\hat{\textbf{p}} = (\cos\theta_c,\sin\theta_c)$ with $\theta_c$ the filament orientation. The force free and torque free conditions are,
%%%
\begin{equation}\label{eq.a7}
\int_{-L/2}^{L/2}\textbf{f}\,ds=0, \quad \int_{-L/2}^{L/2}s\hat{\textbf{p}}\times\textbf{f}\,ds=0.
\end{equation}
%%%
Given the stream function $\Phi$ of the background flow $\textbf{U}$, substitute the filament force [Eq.~(\ref{eq.a2})] into Eq.~(\ref{eq.a7}), we obtain, 
%%%
\begin{equation}\label{eq.a8}
\begin{aligned}
\dot{\textbf{r}}_c &= -\nabla^{\perp}_{\textbf{r}_c}\mathcal{H},\ \text{with}\ \mathcal{H} = \frac{1}{L}\int_{-L/2}^{L/2}\Phi[\textbf{r}(s)]\,ds, \\
\dot{\theta}_c &= \frac{12}{L^2}\mathcal{H}-\frac{6}{L^2}(\Phi[\textbf{r}(L/2)]+\Phi[\textbf{r}(-L/2)]).
\end{aligned}
\end{equation}
%%%
%%%%%%
\begin{figure*}[t]
\centering
\includegraphics[bb=0 15 385 200, scale=0.65,draft=false]{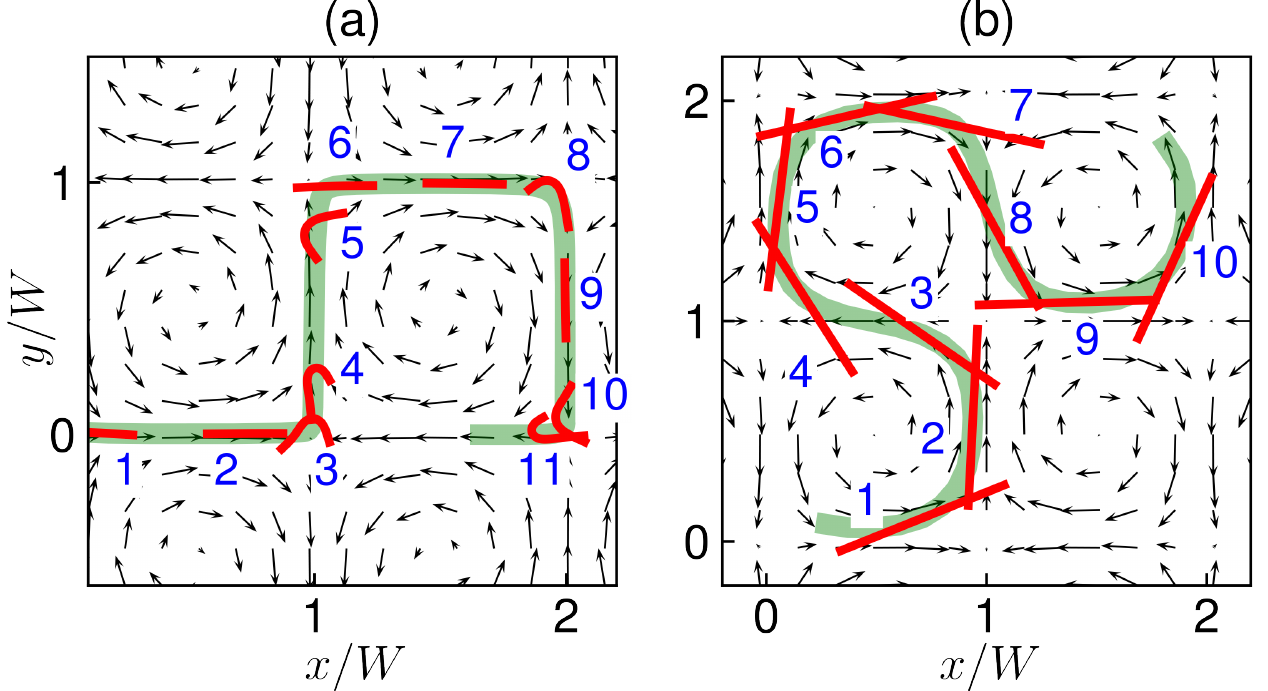}
\caption{From simulations, examples of cross-vortex transport of filaments in a spatially periodic cellular flow. Time ordering is labeled by numbers. The green thick lines trace the CoM trajectories. (a) Short and flexible filament with $\gamma = 0.3$ and $\eta = 1000$. The filament buckles when approaching the stagnation points. (b) Long and rigid filament with $\gamma = 0.8$.}
\label{fig_a2}
\end{figure*}
%%%%

In Fig.~\ref{fig_a2}, we show two examples of flexible and rigid filaments moving across vortices. In Fig.~\ref{fig_a2}(a), the filament is short and flexible. Due to the buckling instabilities at the stagnation points, the filament is deformed and randomly samples the nearby flow regions. Figure~\ref{fig_a2}(b) shows a different mechanism for long and rigid filament: the filament can easily extend across vortices and sample a larger region of the background flow, leading to cross-vortex motion and different types of random walks.

%%%%%%%%%%%%%%%%%%%%%%%%%%%
\section{N\lowercase{umerical} M\lowercase{ethods}}
%%%%%%%%%%%%%%%%%%%%%%%%%%%

We simulate the filament dynamics by solving Eqs.~(\ref{eq.a3})--(\ref{eq.a5}) (flexible filaments) and Eq.~(\ref{eq.a8}) (rigid filaments) using a finite difference method (see Ref.~\cite{{Tornberg}}). We denote by a superscript $n$ quantities at time $t_n$ and discretize the arc length uniformly as $s_j = j \Delta s$, $j=0,\cdots,N$. In simulations of flexible filaments, given the filament configuration ($\textbf{r}^n, \theta^n$) at $t=t_n$, we first solve Eq.~(\ref{eq.a4}) for $T^n$. The spatial derivatives are approximated using a second-order centered difference scheme, while at the boundaries, skew operators are used. To avoid a strict fourth-order stability limit for the time-step size, arising from the fourth-order derivative in Eq.~(\ref{eq.a5}), we use a second-order backward differentiation formula to approximate the time derivative and treat $\theta_{ssss}$ implicitly. All other terms are extrapolated from previous time step. Schematically, we write,
%%%
\begin{equation}\label{eq}
\frac{1}{2\Delta t}(3\theta^{n+1}-4\theta^n+\theta^{n-1}) + \eta^{-1}\theta_{ssss}^{n+1} = 2g[\theta^{n},T^{n}]-g[\theta^{n-1},T^{n-1}],
\end{equation}
%%%
where $g$ denotes the terms on the r.h.s. of Eq.~(\ref{eq.a5}) excluding the fourth-order derivative. We use an Euler scheme for the first time step. For rigid filaments, we evolve Eq.~(\ref{eq.a8}) forward in time using a fourth-order Runge-Kutta scheme (RK4) and the integral along the arc length is approximated using a Simpson's rule. In most of our simulations, $\Delta s = 10^{-2}$ and $\Delta t = 10^{-3}$--$10^{-2}$.

%%%%%%%%%%%%%%%%%%%%%%%%%%%
\section{L\lowercase{yapunov} E\lowercase{xponents} \lowercase{and} P\lowercase{oincar\'e} S\lowercase{ections}}
%%%%%%%%%%%%%%%%%%%%%%%%%%%

Lyapunov exponents characterize the exponential rate at which neighboring phase space trajectories separate on average. We compute the Lyapunov exponents of the dynamical system of rigid filaments using the algorithm in Refs.~\cite{Benettin, Ramasubramanian} based on Gram–Schmidt Reorthonormalizaton (GSR) of the tangent vectors. Equation~(\ref{eq.a8}) is a three dimensional system and can be written as $d\textbf{Z}/dt = \textbf{F}(\textbf{Z},t)$, where $\textbf{Z}$ is three dimensional, $\textbf{Z} = [\textbf{r}_c, \theta_c]$, and $\textbf{F}$ is also a three dimension vector from the r.h.s. of Eq.~(\ref{eq.a8}). Define the Jacobian matrix $\textbf{J}$ with $J_{ij} = \partial F_i /\partial Z_j$, we summarize the algorithm below,
%%%%%%%%%%%%%%%%%%%%%%
\begin{algorithmic}[1]
\State $t = 0$, $i = 1$, $\textbf{Z}$ = [$\textbf{r}_c^0$, $\theta_c^0$], and $\textbf{A} = [\hat{\textbf{e}}_1, \hat{\textbf{e}}_2, \hat{\textbf{e}}_3]$, where $\hat{\textbf{e}}_1$, $\hat{\textbf{e}}_2$, and $\hat{\textbf{e}}_3$ are orthogonal vectors;    \Comment{Initialization}
\State Initialize a $J\times3$ matrix $\textbf{M}$;   
\While {$t<=J\tau$}
    \State $\textbf{Z} \gets \texttt{RK4}(\textbf{Z}, \textbf{F})$; $\textbf{A} \gets \texttt{RK4}(\textbf{A}, \textbf{J}\cdot \textbf{A})$;    \Comment{Evolve $d\textbf{Z}/dt = \textbf{F}$ and $d\textbf{A}/dt = \textbf{J}\cdot\textbf{A}$}
    
    \If {$\texttt{rem}(t,\tau) = 0$}    \Comment{At time $t = k\tau$ with k positive integers}
        \State $\textbf{A}, \textbf{M}_{i,:} \gets \texttt{GSR}(\textbf{A})$;  \Comment{GSR of the columns of $\textbf{A}$}
        \State $i \gets i+1$;
    \EndIf
    \State $t \gets t+\Delta t$; \Comment{Update time step}
\EndWhile
\State $\lambda_j = \sum_{i=1}^{J} \log(M_{ij})/(J\tau)$;    \Comment{Compute Lyapunov exponents}
\end{algorithmic}
%%%%%%%%%%%%%%%%%%
Here, \texttt{RK4} implements a fourth-order Runge-Kutta scheme, which takes a variable and its derivative function as inputs and update the variable one time step forward. \texttt{GSR} implements Gram–Schmidt orthonormalization, which returns the orthonormalized matrix and the norms of the orthogonalized vectors. The norms are stored on the $i$-th row of $\textbf{M}$ and used to compute the Lyapunov exponents. In Table~\ref{table_a2}, we show several results along with the parameters used from our implements. Figure 3(a) in the main text shows the largest Lyapunov exponent ($\lambda_1$) averaged from an ensemble of random initial conditions.
%%%%%%%%%%%%%%%%%%%
\begin{table}[h]
    \centering
    \begin{tabular}{p{1.5cm}|p{1cm}p{1cm}p{1cm}p{1cm}p{1cm}p{1.3cm}|p{1.5cm}p{1.5cm}p{1.5cm}}
           & $x_c^0/W$ & $y_c^0/W$ & $\theta_c^0$ & $\Delta t$ & $J$ & $\tau (L/U_0)$ & $\lambda_1$ & $\lambda_2$ & $\lambda_3$ \\
        \hline
        $\gamma = 0.7$ & 0.1 & 0.1 & $\pi/6$ & 0.01 & 1070 & 0.36 & 0.22032 & -0.00035 & -0.23374 \\
        \hline
        $\gamma = 1.0$ & 0.1 & 0.1 & $\pi/6$ & 0.01 & 750 & 0.25 & 0.39158 & 0.00045 & -0.42781 \\
        \hline
        $\gamma = 1.4$ & 0.1 & 0.1 & $\pi/6$ & 0.01 & 540 & 0.18 & 0.71167 & 0.00051 & -0.73723
    \end{tabular}
    \caption{From simulations, Lyapunov exponents for three different values of $\gamma$ and the parameters used.}
    \label{table_a2}
\end{table}
%%%%%%%%%%%%%%%%%%
%%%
\begin{figure*}[b]
\centering
\includegraphics[bb=0 5 365 90, scale = 1.3,draft=false]{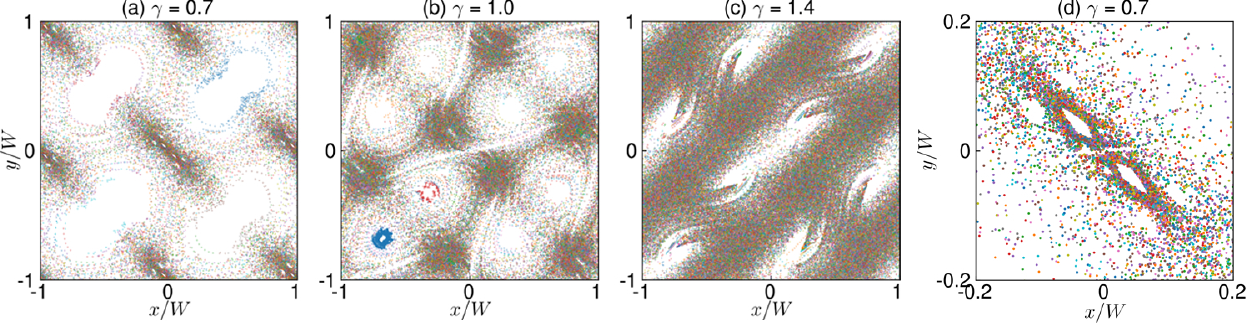}
\caption{Poincar\'e maps constructed by recording the center of mass (CoM) coordinates $(x_c, y_c)$ when $\theta_c = \pi/4$ and then mapping to the fundamental unit cell. (a) $\gamma = 0.7$, (b) $\gamma = 1.0$, and (c) $\gamma = 1.4$. (d) A zoom-in view of the region $[-0.2, 0.2] \times [-0.2,0.2]$ for $\gamma = 0.7$. The filaments are initially located outside the islands. Different colors indicate trajectories with different initial conditions.}
\label{fig_a3}
\end{figure*}
%%%

\begin{figure*}[t]   
\centering
\includegraphics[bb=0 8 260 173, scale = 0.55,draft=false]{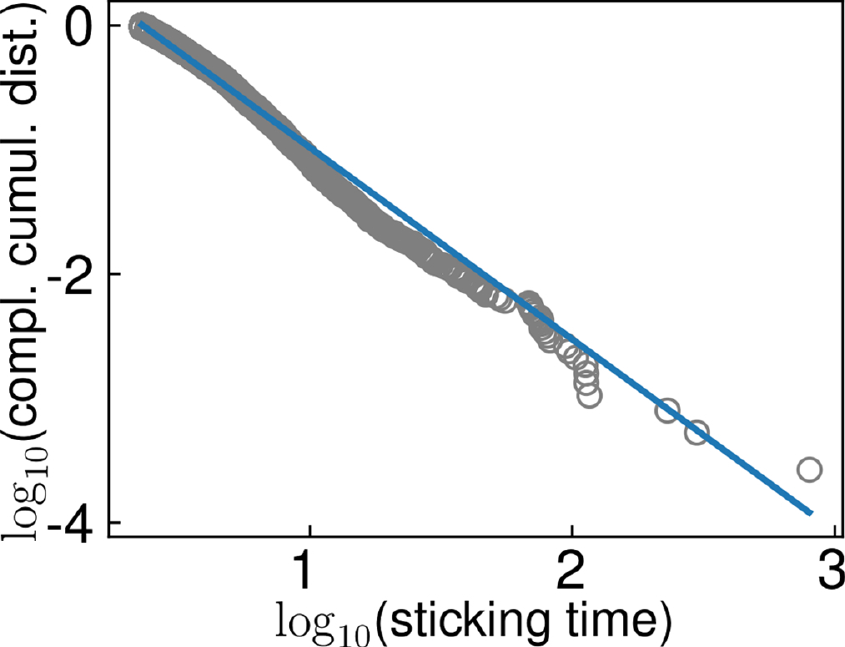}
\caption{Distribution of the sticking times in the region $[-0.1, 0.1]\times [-0.1, 0.1]$ for $\gamma = 0.7$. The blue line is the best-fit power law to the distribution.}
\label{fig_a4}
\end{figure*}

For rigid filaments, it can be shown that Eqs.~(\ref{eq.a8}) has the following reversal symmetry: $(t, \textbf{Z}) \mapsto (-t, G \textbf{Z})$, where $G$ is the phase space involution, $G: (x_c, y_c, \theta_c) \mapsto (y_c, x_c, \pi/2-\theta_c)$. This indicates that our system may show properties similar to Hamiltonian systems, there may also exist structures associated with dissipative systems~\cite{Roberts, Ariel20}. In Fig.~\ref{fig_a3}, we show Poincar\'e maps for different values of $\gamma$. When $\gamma = 0.7$, we observe strong nonuniformity in the phase space. In particular, a higher density of points are accumulated in the regions around regular islands, indicating that these regions are frequently visited by the filaments. In Fig.~\ref{fig_a3}(d), a zoom-in view of the region $[-0.2, 0.2] \times [-0.2, 0.2]$ shows trajectories sticking to 6 islands (two 'figure of 8' islands in the center and four islands surrounding them). In this work, we have not explored the detailed properties of these islands, i.e., periodic orbits and their stability. This stickiness in phase space leads to long persistent steps in L\'evy walks. We compute the distribution of the sticking times. In our study, similar to Ref.~\cite{Ariel20}, we define the sticking time as the time the trajectories spend within the small square regions $[-0.1,0.1] \times [-0.1,0.1]$ surrounding the regular islands. Figure~\ref{fig_a4} shows that the complement of the cumulative distribution of the sticking times follows approximately a power law. When $\gamma = 1.0$ [see Fig.~\ref{fig_a3}(b)], the sticking regions becomes larger and less concentrated. More points are spread over the phase space. When $\gamma = 1.4$, no islands and sticking regions are observed and most of phase space are occupied. On the other hand, in Fig.~\ref{fig_a3}(a), the islands around the center of the vortices account for occasional trapping events near the center of the vortices. These regions becomes smaller as $\gamma$ increases. The above discussion presents a view of the general structure of the phase space and its dependence on $\gamma$. We may focus on the finer structures in our future work.
%%%
%%%

%%%%%%%%%%%%%%%%%%%%%%%%%%%
\section{S\lowercase{tep-length} D\lowercase{istribution} A\lowercase{nalysis}}
%%%%%%%%%%%%%%%%%%%%%%%%%%%

\begin{figure*}[t]
\centering
\includegraphics[bb=0 0 265 200, scale=0.6,draft=false]{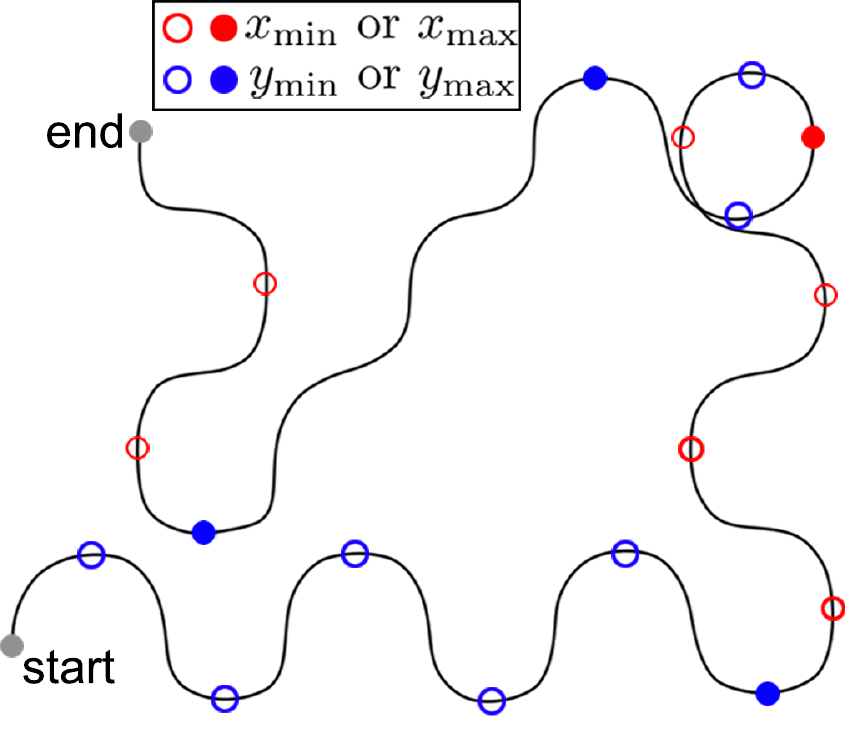}
\caption{Determination of the turning points by finding the local minimums and maximums of the $x$ and $y$ coordinates of the CoM trajectory. The red circles are the extremums of x coordinates and the blue circles are the extremums of y coordinates. Open circles are discarded and filled circles are determined turning points (including the start and end points).}
\label{fig_a5}
\end{figure*}

The step length $l$ is the straight-line distance between successive turning points separating steps along different directions. To identify the turning points, we first find the local minimums and maximums of the $x$ coordinate of the CoM trajectories [Fig.~\ref{fig_a5}], and then calculate the straight-line distances between each two neighbor extremums. If a distance is smaller than a minimum value $l_{\text{min}}$, we discard the corresponding leading extremum used in computing that distance. The same procedure is repeated for $y$ coordinates. The turning points are the aggregate of the remaining extremums of $x$ and $y$ coordinates while keeping their time ordering. We take $l_{\text{min}}$ to be slightly larger than the vortex size $W$.

\begin{figure*}[t]   
\centering
\includegraphics[bb=0 5 365 80, scale = 1.3,draft=false]{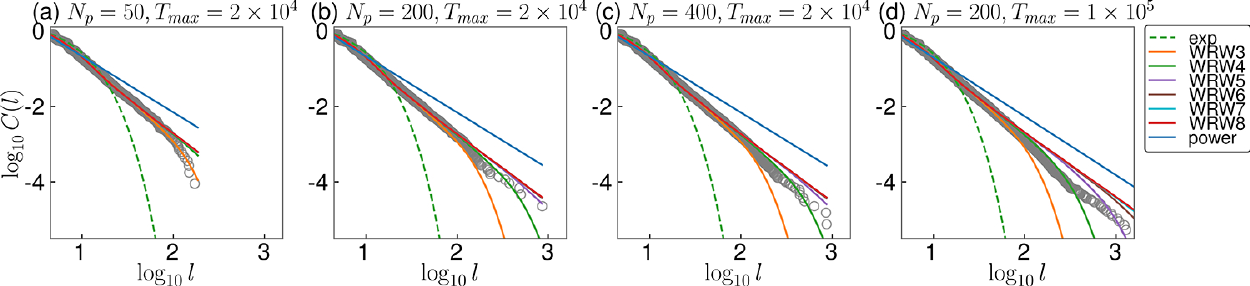}
\caption{From simulations, step length distribution $\phi(l)$ and several best-fit random walk models for $l$ obtained from different number of trajectories $N_p$ and total simulation time $T_{max}$. The filament relative length $\gamma = 1.0$.}
\label{fig_a6}
\end{figure*}

After we obtain the step-length data, we fit it to the candidate distributions mentioned in the main text using a maximum likelihood method~\cite{David,Clauset,Edwards}. The Akaike weight $w$ is then calculated for model selection~\cite{Anderson}. The data completely supports the distribution when $w=1$ and does not support the distribution when $w=0$. The above approach is robust for identifying L\'evy walk patterns. We summarize the expressions for the candidate distributions in Eq.~(\ref{eq.a11}),

For WRWs, the highest order we have considered is $J = 8$. In the simulations, we observe that the Akaike weights $w$ of higher-tier WRWs increase as we increase the number of statistics. An example is shown in Fig.~\ref{fig_a6} with $\gamma = 1.0$. First, we increase the number of trajectories $N_p$ used to compute the step length distribution [Figs.~\ref{fig_a6}(a)--(c)]. When $N_p = 50$, the best fit model is a 3-tier WRW. However, when $N_p = 200$ [Fig.~\ref{fig_a6}(b)], with more trajectories, longer steps appears more frequently at the tail of the distribution. The best-fit models are 4-tier and 5-tier WRWs. As $N_p$ further increases [Fig.~\ref{fig_a6}(c)], the distribution does not improve much and the convergence to higher-tier distributions seems to be slow. Since the steps drawn from higher tier distributions are less frequent, to see higher-tier distributions, we increase the total simulation time $T_{max}$ from $2\times 10^4 (L/U_0)$ to $1\times 10^5 (L/U_0)$ in Fig.~\ref{fig_a6}(d). Although the bulk fits to a 4-tier distribution, the tail fits better to a 5-tier distribution with longer steps becomes more noticeable, and the distribution looks closer to a power law. 

In the experiments, since the flow field is bounded, we fit the step-length data to the truncated power law and truncated exponential distribution. The Akaike weights for $\gamma = 0.82$, 1.26, and 1.5 with different values of $\eta$ are shown in Table~\ref{table_a3}. The step-length data for short filaments ($\gamma = 0.82$ and 1.26) better supports truncated power law. For longer filament ($\gamma = 1.5$), the step-length data better supports truncated exponential distribution.
%%%
%%%
%%%
%%%
%%%
\begin{equation}\label{eq.a11}
\begin{aligned}
\text{power law: } \phi(l) &= \beta l_{min}^{\beta} l^{-(\beta+1)},\ &l_{min} \le l, \\
\text{truncated power law: } \phi(l) &= \frac{\beta l^{-(\beta+1)}}{l_{min}^{-\beta}-l_{max}^{-\beta}},\ &l_{min}\le l \le l_{max}, \\
\text{exponential distribution: } \phi(l) &= b \exp(-b(l-l_{min})),\ &l_{min} \le l, \\
\text{truncated exponential distribution: } \phi(l) &= \frac{b \exp(-b(l-l_{min}))}{1-\exp(-b(l_{max}-l_{min}))},\ &l_{min} \le l \le l_{max},\\
\text{Weierstrassian random walk: } \phi(l) &= (q-1)\sum_{j=0}^{J} q^{-(j+1)} b^{j+1}\exp(-b^{j+1} (l-l_{min})),\ &l_{min} \le l.
\end{aligned}
\end{equation}
%%%
%%%
\begin{table}[h]
    \centering
    \begin{tabular}{ p{1.5cm}|p{1cm}p{3.5cm}p{2.4cm} }
    
            & $\eta$ & Distribution & Akaike weights $w$ \\
        \hline
        \multirow{2}{*}{$\gamma = 0.82$} & \multirow{2}{*}{25} & truncated exponential & 0.05 \\
            &   & truncated power law & 0.95 \\
        \hline
        \multirow{6}{*}{$\gamma = 1.26$} & \multirow{2}{*}{35} & truncated exponential & 0 \\
            &   & truncated power law & 1 \\
        \cline{2-4}
            &   \multirow{2}{*}{85} & truncated exponential & 0 \\
            &   & truncated power law & 1 \\
        \cline{2-4}
            &   \multirow{2}{*}{112} & truncated exponential & 0.04 \\
            &   & truncated power law & 0.96 \\
        \hline
        \multirow{4}{*}{$\gamma = 1.50$} & \multirow{2}{*}{118} & truncated exponential & 1 \\
            &   & truncated power & 0 \\
        \cline{2-4}
            &   \multirow{2}{*}{166} & truncated power law & 1 \\
            &   & truncated power law & 0 \\
        \hline
    \end{tabular}
    \caption{From experiments, Akaike weights of best-fit truncated power law and truncated exponential distribution for $\gamma = 0.82, 1.26$, and 1.5 with different values of $\eta$.}
    \label{table_a3}
\end{table}
%%%
%%%%%%%%%%%%%%%%%%%%%%%%%%%%%%%%%%%%%%%%%%%%%%%%%
\section{D\lowercase{ynamical} S\lowercase{orting} \lowercase{in the} B\lowercase{allistic} S\lowercase{tate}}
%%%%%%%%%%%%%%%%%%%%%%%%%%%%%%%%%%%%%%%%%%%%%%%%%
%%%%%%
\begin{figure*}[h]
\centering
\includegraphics[bb=0 0 350 100, scale=1.2,draft=false]{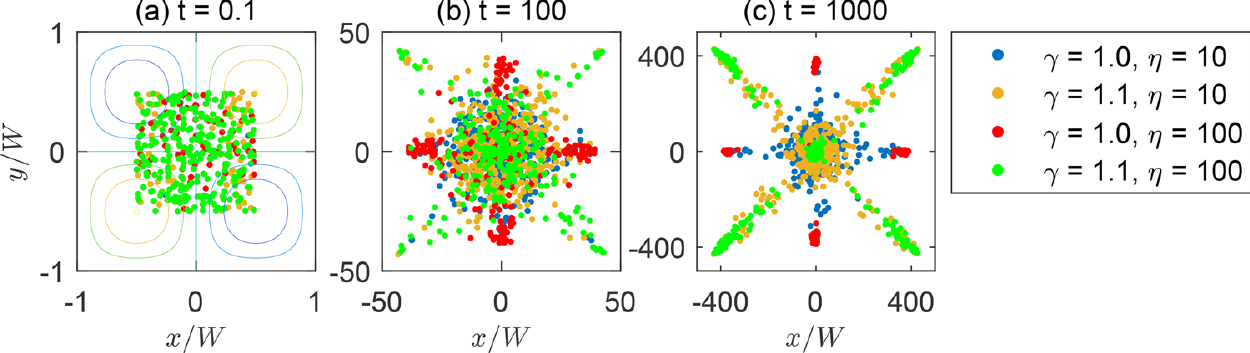}
\caption{From simulation, anisotropic dispersals in ballistic state showing the distribution of the CoM positions for four different filament lengths and flexibility: $\gamma = 1, \eta = 10$ (blue), $\gamma = 1.1, \eta = 10$ (yellow), $\gamma = 1, \eta = 100$ (red), and $\gamma = 1.1, \eta = 100$ (green), at (a) $t=0.1$, (b) $t=100$, and (c) $t=1000$.}
\label{fig_a7}
\end{figure*}
%%%%
In the ballistic state, the filament typically meanders around for a short period $t_m$ before taking steps along diagonals or $\pm x$ and $\pm y$ directions. Longer filaments ($\gamma > 1$) are transported along diagonals and shorter filaments ($\gamma \le 1$) are transported along $\pm x$ and $\pm y$ directions. The dispersals of filaments is anisotropic and depend strongly on $\gamma$ and $\eta$~[Fig.~\ref{fig_a7}]. Initially~[Fig.~\ref{fig_a7}(a)], four different filaments with different values of $\gamma$ and $\eta$ are distributed inside a square region ($x \in [-W/2, W/2]$, $y \in [-W/2, W/2]$) with random initial positions and orientations. The anisotropy in the dispersal patterns grows quickly. When $t=100$~[Fig.~\ref{fig_a7}(b)], some of the green ($\gamma = 1.1, \eta = 100$) and red filaments ($\gamma = 1, \eta = 100$) are already separated spatially with the green filaments moving along diagonals and red filaments moving along $\pm x$ and $\pm y$ directions. When $t$ is sufficiently large [Fig.~\ref{fig_a7}(c)], four kinds of filaments are already well separated from each other. Filaments with larger $\eta$ mainly occupy the dispersal fronts, and filaments with smaller $\eta$ are distributed along diagonals or $x$ and $y$ directions, i.e., the meandering time $t_m$ decreases as $\eta$ increases. However, filaments with sufficiently large $\eta$ will result in trapping state, in which filaments are bent with large deformation and trapped inside one of the vortices.

%%%%%%%%%%%%%%%%%%%%%%%%%%%%%%%%%%%%%%%%%%%%%%%%
\section{P\lowercase{erturbation} \lowercase{of the} B\lowercase{ackground} F\lowercase{low}}
%%%%%%%%%%%%%%%%%%%%%%%%%%%%%%%%%%%%%%%%%%%%%%%%
%%%%%%
\begin{figure*}[h]
\centering
\includegraphics[bb=0 0 350 120, scale=1.,draft=false]{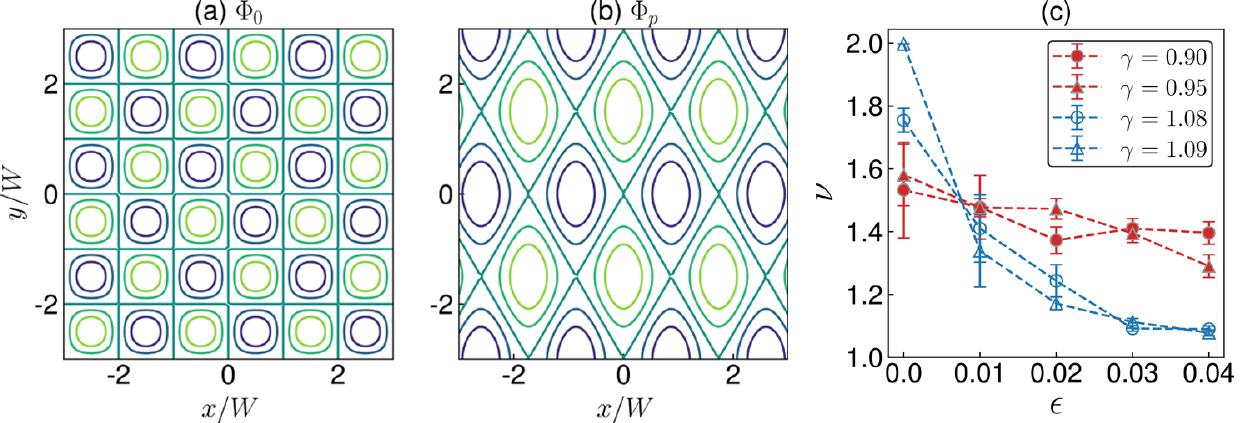}
\caption{(a) Unperturbed stream function. (b) Perturbation stream function $\Phi_p$ given by Eq.~(\ref{eq.a11}) with $\xi = 2/3$, $\alpha = 1/2$ and $\beta = \sqrt{3}/2$. (c) The scaling exponent $\nu$ of the mean square displacement as function of $\epsilon$ for different values of $\gamma$.}
\label{fig_a8}
\end{figure*}
%%%%
For $\gamma$ around 1.08--1.09, we observe nearly ballistic motion with $\nu \approx 2$ [see Fig. 2(b) main text]. This is due to a geometric match of the filament length to the periodic flow field, which can be verified by perturbing the background flow. The unperturbed stream function $\Phi_0 = (\pi\gamma)^{-1}\sin(\pi\gamma x)\sin(\pi\gamma y)$. The perturbation stream function $\Phi_p$ is given by~\cite{Ariel17}.
\begin{equation}\label{eq.a12}
\Phi_{p} = (\pi \gamma)^{-1}\sin\left[\pi\gamma \xi(\alpha x+\beta y)\right]\sin\left[\pi\gamma \xi (\alpha x-\beta y)\right],
\end{equation}
where $\xi$ controls the characteristic scale of the flow, $\alpha$ and $\beta$ change the direction of the separatrixes [Fig.~\ref{fig_a8}(b)]. The resulting flow field is given by $\Phi = \Phi_0+\epsilon \Phi_p$, where $\epsilon$ is the perturbation magnitude. We take $\xi = 2/3$, $\alpha = 1/2$ and $\beta = \sqrt{3}/2$, and $\Phi_p$ is still spatially periodic but the period is different from that of $\Phi_0$. Figure~\ref{fig_a8}(c) shows that for $\gamma = 1.08$ and $\gamma = 1.09$, the scaling exponent of the mean square displacement, $\nu$, decreases rapidly as $\epsilon$ increases. However, for $\gamma = 0.9$ and 0.95, $\nu$ is relatively stable and decreases much slower.

%%%%%%%%%%%%%%%%%%%%%%%%%%

%%%%%%%%%%%%%%%%%%%%%%%%%%%%%%%%%